\documentclass[11pt]{article}
\usepackage{amsmath,amsfonts,amsthm,times,latexsym,graphicx}

\newtheoremstyle{localthm}
	{5pt} 
	{5pt} 
	{\sl} 
	{} 
	{\bf} 
	{} 
	{.7em} 
	{} 

\newtheoremstyle{localrem}
	{5pt} 
	{5pt} 
	{\rm} 
	{} 
	{\bf} 
	{} 
	{.7em} 
	{} 

\theoremstyle{localthm}
\newtheorem{Theorem}{Theorem}[section]

\theoremstyle{localrem}
\newtheorem{Example}[Theorem]{Example}

\newcommand{\ruck}[1]{\strut\hspace{#1cm}}
\newcommand{\dom}{\mathrm{dom}}

\def\bs{\boldsymbol}

\def\R{\mathbb{R}}

\def\eps{\epsilon}

\def\GG{\mathcal{G}}
\def\KK{\mathcal{K}}
\def\VV{\mathcal{V}}

\def\til{\widetilde}
\def\hat{\widehat}

\def\Ex{\mathop{\rm I\!E}\nolimits}

\def\Var{\mathrm{Var}}

\newcommand{\GGconc}{\GG_{\mathrm{conc}}}

\begin{document}

\addtolength{\baselineskip}{+.5\baselineskip}

\title{Active Set and EM Algorithms for Log-Concave Densities Based on Complete and Censored Data}
\author{Lutz D\"{u}mbgen, Andr\'{e} H\"{u}sler and Kaspar Rufibach\\
	University of Bern}

\date{August 2007, revised March 2011}

\maketitle

\textbf{Abstract.}
We develop an active set algorithm for the maximum likelihood estimation of a log-concave density based on complete data. Building on this fast algorithm, we indicate an EM algorithm to treat arbitrarily censored or binned data.

\section{Introduction}
\label{Introduction}

A probability density $f$ on the real line is called log-concave if it may be written as
$$
	f(x) \ = \ \exp \phi(x)
$$
for some concave function $\phi : \R \to [-\infty,\infty)$. The class of all log-concave densities provides an interesting nonparametric model consisting of unimodal densities and containing many standard parametric families; see D\"umbgen and Rufibach (2009) for a more thorough overview.

This paper treats algorithmic aspects of maximum likelihood estimation for this particular class. In Section~\ref{Complete Data} we derive a general finite-dimensional optimization problem which is closely related to computing the maximum likelihood estimator of a log-concave probability density $f$ based on independent, identically distributed observations. Section~\ref{Active Set} is devoted to the latter optimization problem. At first we describe generally an active set algorithm, a useful tool from optimization theory (cf.\ Fletcher, 1987) with many potential applications in statistical computing. A key property of such algorithms is that they terminate after finitely many steps (in principle). Then we adapt this approach to our particular estimation problem, which yields an alternative to the iterative algorithms developed by Rufibach (2006, 2007) and Pal, Woodroofe and Meyer (2006). The resulting active set algorithm is similar in spirit to the vertex direction and support reduction algorithms described by Groeneboom, Jongbloed and Wellner (2008), who consider the special setting of mixture models.

In Section~\ref{Censored Data} we consider briefly the problem of estimating a probability distribution $P$ on $(0,\infty]$ based on censored or binned data. Censoring occurs quite frequently in biomedical applications, e.g.\ $X$ being the time point when a person develops a certain disease or dies from a certain cause. Another field of application is quality control where $X$ is the failure time of a certain object. A good reference for event time analysis is the monograph of Klein and Moeschberger (1997). Binning is typical in socioeconomic surveys, e.g.\ when persons or households are asked which of several given intervals their yearly income $X$ falls into. We discuss maximum likelihood estimation of $P$ under the assumption that it is absolutely continuous on $(0,\infty)$ with log-concave probability density $f$. The resulting estimator is an alternative to those of D\"{u}mbgen et al.\ (2006). The latter authors restrict themselves to interval-censored data and considered the weaker constraints of $f$ being non-increasing or unimodal. Introducing the stronger but still natural constraint of log-concavity allows us to treat arbitrarily censored data, similarly as Turnbull (1976). In Section~\ref{EM} we indicate an expectation-maximization (EM) algorithm for the estimation of $P$, using the aforementioned active set algorithm as a building block. This approach is similar to Turnbull (1976) and Braun et al.\ (2005); the latter authors considered self-consistent kernel density estimators. For more information and references on EM and related algorithms in general we refer to Lange et al.\ (2000). A detailed description of our method for censored or binned data will be given elsewhere.

Section~\ref{Proofs} contains most proofs and various auxiliary results.

\section{The general log-likelihood function for complete data}
\label{Complete Data}

\paragraph{Independent, identically distributed observations.}
Let $X_1, X_2, \ldots, X_n$ be independent random variables with log-concave probability density $f = \exp\phi$ on $\R$. Then the normalized log-likelihood function is given by
$$
	\ell(\phi) \ := \ n_{}^{-1} \sum_{i=1}^n \phi(X_i) .
$$
It may happen that due to rounding errors one observes $\til{X}_i$ in place of $X_i$. In that case, let $x_1 < x_2 < \cdots < x_m$ be the different elements of $\{\til{X}_1, \til{X}_2,\ldots,\til{X}_n\}$ and define $p_i := n^{-1} \# \{j : \til{X}_j = x_i\}$. Then an appropriate surrogate for the normalized log-likelihood is
\begin{equation}
	\ell(\phi) \ := \ \sum_{i=1}^m p_i \phi(x_i) .
	\label{eq: log-lik}
\end{equation}

\paragraph{The general log-likelihood function.}
In what follows we consider the functional \eqref{eq: log-lik} for arbitrary given points $x_1 < x_2 < \cdots < x_m$ and probability weights $p_1, p_2, \ldots, p_m > 0$, i.e.\ $\sum_{i=1}^m p_i = 1$. Suppose that we want to maximize $\ell(\phi)$ over all functions $\phi$ within a certain family $\mathcal{F}$ of measurable functions from $\R$ into $[-\infty,\infty)$ satisfying the constraint $\int \exp \phi(x) \, dx = 1$. If $\mathcal{F}$ is closed under addition of constants, i.e.\ $\phi + c \in \mathcal{F}$ for arbitrary $\phi \in \mathcal{F}$ and $c \in \R$, then one can easily show that maximizing $\ell(\phi)$ over all $\phi \in \mathcal{F}$ with $\int \exp \phi(x) \, dx = 1$ is equivalent to maximizing
$$
	L(\phi) \ := \ \sum_{i=1}^m p_i \phi(x_i) - \int \exp \phi(x) \, dx
$$
over the whole family $\mathcal{F}$; see also Silverman~(1982, Theorem~3.1).

\paragraph{Restricting the set of candidate functions.}
The preceding considerations apply in particular to the family $\mathcal{F}$ of all concave functions. Now let $\GG$ be the set of all continuous functions $\psi : [x_1,x_m] \to \R$ which are linear on each interval $[x_k, x_{k+1}]$, $1 \le k < m$, and we define $\psi := - \infty$ on $\R \setminus [x_1,x_m]$. Moreover, let $\GGconc$ be the set of all concave functions within $\GG$. For any $\phi \in \mathcal{F}$ with $L(\phi) > - \infty$ let $\psi$ be the unique function in $\GGconc$ such that $\psi = \phi$ on $\{x_1,x_2,\ldots,x_m\}$. Then it follows from concavity of $\phi$ that $\psi \le \phi$ pointwise, and $L(\psi) \ge L(\phi)$. Equality holds if, and only if, $\psi = \phi$. Thus maximizing $L$ over the class $\mathcal{F}$ is equivalent to its maximization over $\GGconc$.

\paragraph{Properties of $L(\cdot)$.}
For explicit calculations it is useful to rewrite $L(\psi)$ as follows: Any function $\psi \in \GG$ may be identified with the vector $\bs{\psi} := (\psi(x_i))_{i=1}^m \in \R^m$. Likewise, any vector $\bs{\psi} \in \R^m$ defines a function $\psi \in \GG$ via
$$
	\psi(x) \ := \ \Bigl( 1 - \frac{x - x_k}{\delta_k} \Bigr) \, \psi_k^{}
		+ \frac{x - x_k}{\delta_k} \, \psi_{k+1}^{}
	\quad\mbox{for } x \in [x_k,x_{k+1}], 1 \le k < m ,
$$
where $\delta_k := x_{k+1} - x_k$. Then one may write
$$
	L(\psi) \ = \ L(\bs{\psi}) := \sum_{i=1}^m p_i \psi_i - \sum_{k=1}^{m-1} \delta_k J(\psi_k, \psi_{k+1})
$$
with
$$
	J(r,s) \ := \ \int_0^1 \exp \bigl( (1 - t) r + t s \bigr) \, dt
$$
for arbitrary $r,s \in \R$. The latter function $J : \R \times \R \to \R$ is infinitely often differentiable and strictly convex. Hence $L(\cdot)$ is an infinitely often differentiable and strictly concave functional on $\R^m$. In addition it is coercive in the sense that
\begin{equation}
	L(\bs{\psi}) \ \to \ -\infty
	\quad\mbox{as } \|\bs{\psi}\| \to \infty .
	\label{eq: coercivity of L}
\end{equation}
This entails that both
\begin{eqnarray}
	\til{\psi} & := & \mathop{\rm argmax}_{\psi \in \GG} L(\psi)
	\label{eq: definition of psicheck}
	\quad\mbox{and} \\
	\hat{\psi} & := & \mathop{\rm argmax}_{\psi \in \GGconc} L(\psi)
	\label{eq: definition of psihat}
\end{eqnarray}
are well defined and unique.

Let us discuss some further properties of $L(\cdot)$ and its unrestricted maximizer $\til{\psi}$. To maximize $L(\cdot)$ we need its Taylor expansion of second order. In fact, for functions $\psi, v \in \GG$,
\begin{eqnarray}
	\frac{d}{dt} \Big\vert_{t=0} L(\psi + tv)
	& = & \sum_{i=1}^m p_i v(x_i) - \int v(x) \exp \psi(x) \, dx ,
	\label{eq: 1st dir deriv L} \\
	\frac{d^2}{dt^2} \Big\vert_{t=0} L(\psi + tv)
	& = & - \int v(x)^2 \exp \psi(x) \, dx .
	\label{eq: 2nd dir deriv L}
\end{eqnarray}
Note that the latter expression yields an alternative proof of $L$'s strict concavity. Explicit formulae for the gradient and hessian matrix of $L$ as a functional on $\R^m$ are given in Section~\ref{Proofs}, and with these tools one can easily compute $\til{\psi}$ very precisely via Newton type algorithms. We end this section with a characterization and interesting properties of the maximizer $\til{\psi}$. In what follows let
$$
	J_{ab}(r,s) \ := \ \frac{\partial^{a+b}}{\partial r^a \partial s^b} J(r,s)
	\ = \ \int_0^1 (1 - t)^a t^b \exp((1 - t)r + t s) \, dt .
$$
for nonnegative integers $a$ and $b$.

\begin{Theorem}
\label{thm: 1dim functional}
Let $\psi \in \GG$ with corresponding density $f(x) := \exp \psi(x)$ and distribution function $F(r) := \int_{x_1}^r f(x) \, dx$ on $[x_1, x_m]$. The function $\psi$ maximizes $L$ if, and only if, its distribution function $F$ satisfies
$$
	F(x_m) = 1
	\quad\mbox{and}\quad
	\delta_k^{-1} \int_{x_k}^{x_{k+1}} F(x) \, dx
	\ = \ \sum_{i=1}^k p_i
	\quad\mbox{for } 1 \le k < m .
$$
In that case,
$$
	\int_{x_1}^{x_m} x f(x) \, dx \ = \ \sum_{i=1}^m p_i x_i
$$
and
$$
	\int_{x_1}^{x_m} x^2 f(x) \, dx
	\ = \ \sum_{i=1}^m p_i^{} x_i^2
		- \sum_{k=1}^{m-1} \delta_k^3 J_{11}(\psi_k, \psi_{k+1}) .
$$
\end{Theorem}

\paragraph{Some auxiliary formulae.}
For $\psi \in \GG$ with density $f(x) := \exp \psi(x)$ and distribution function $F(r) := \int_{x_1}^r f(x) \, dx$ on $[x_1, x_m]$, one can easily derive explicit expressions for $F$ and the first two moments of $f$ in terms of $J(\cdot,\cdot)$ and its partial derivatives: For $1 \le k < m$,
$$
	F(x_{k+1}) \ = \ \sum_{i=1}^k \delta_i J(\psi_i, \psi_{i+1})
$$
and
$$
	\delta_k^{-1} \int_{x_k}^{x_{k+1}} F(x) \, dx
	\ = \ F(x_k) + \delta_k J_{10}(\psi_k,\psi_{k+1})
	\ \in \ \bigl( F(x_k), F(x_{k+1}) \bigr) .
$$
Moreover, for any $a \in \R$,
\begin{eqnarray*}
	\int_{x_1}^{x_m} (x-a) f(x) \, dx
	& = & \sum_{k=1}^{m-1} \delta_k \bigl( (x_k - a) J_{10}(\psi_k,\psi_{k+1})
		+ (x_{k+1} - a) J_{01}(\psi_k,\psi_{k+1}) \bigr) , \\
	\int_{x_1}^{x_m} (x - a)^2 f(x) \, dx
	& = & \sum_{k=1}^{m-1} \delta_k \bigl( (x_k - a)^2 J_{10}(\psi_k,\psi_{k+1})
		+ (x_{k+1} - a)^2 J_{01}(\psi_k,\psi_{k+1}) \bigr) \\
	&& - \ \sum_{k=1}^{m-1} \delta_k^3 J_{11}(\psi_k,\psi_{k+1}) .
\end{eqnarray*}

\section{An active set algorithm}
\label{Active Set}

\subsection{The general principle}

We consider an arbitrary continuous and concave function $L : \R^m \to [-\infty,\infty)$ which is coercive in the sense of \eqref{eq: coercivity of L} and continuously differentiable on the set $\dom(L) := \{\bs{\psi} \in \R^m : L(\bs{\psi}) > - \infty\}$. Our goal is to maximize $L$ on the closed convex set
$$
	\KK
	\ := \ \left\{ \bs{\psi} \in \R^m : \bs{v}_i^\top \bs{\psi} \le c_i \ \text{for} \ i = 1,\ldots,q \right\} ,
$$
where $\bs{v}_1,\ldots,\bs{v}_q$ are nonzero vectors in $\R^m$ and $c_1, \ldots, c_q$ real numbers such that $\KK \cap \dom(L) \ne \emptyset$. These assumptions entail that the set
$$
	\KK_* \ := \ \mathop{\rm argmax}_{\bs{\psi} \in \KK} \, L(\bs{\psi})
$$
is a nonvoid and compact subset of $\dom(L)$. For simplicity we shall assume that
\begin{equation}
	\bs{v}_1, \bs{v}_2, \ldots, \bs{v}_q \ \text{are linearly independent} ,
	\label{ass: linear independence}
\end{equation}
but see also the possible extensions indicated at the end of this section.

An essential tacit assumption is that for any index set $A \subseteq \{1,\ldots,q\}$ and the corresponding affine subspace
$$
	\VV(A) \ := \ \left\{ \bs{\psi} \in \R^m : \bs{v}_a^\top \bs{\psi} = c_a \mbox{ for all } a \in A \right\}
$$
of $\R^m$, we have an algorithm computing a point
$$
	\til{\bs{\psi}}(A) \ \in \ \VV_*(A) \ := \ \mathop{\rm argmax}_{\bs{\psi} \in \VV(A)} \, L(\bs{\psi}) ,
$$
provided that $\VV(A) \cap \dom(L) \ne \emptyset$. Now the idea is to vary $A$ suitably until, after finitely many steps, $\til{\bs{\psi}}(A)$ belongs to $\KK_*$.

In what follows we attribute to any vector $\bs{\psi} \in \R^m$ the index set
$$
	A(\bs{\psi}) \ := \ \Bigl\{ i \in \{1,\ldots,q\} \ : \ \bs{v}_i^\top \bs{\psi} \ge c_i \Bigr\} .
$$
For $\bs{\psi} \in \KK$ the set $A(\bs{\psi})$ identifies the ``active constraints'' for $\bs{\psi}$. The following theorem provides useful characterizations of $\KK_*$ and $\VV_*(A)$.

\begin{Theorem}
\label{thm: KKstar and VVA}
Let $\bs{b}_1,\ldots,\bs{b}_m$ be a basis of $\R^m$ such that
$$
	\bs{v}_i^\top \bs{b}_j^{} \ \begin{cases}
		< \ 0 & \text{if} \ i = j \le q , \\
		= \ 0 & \text{else} .
	\end{cases}
$$

\noindent
\textbf{(a)} A vector $\bs{\psi} \in \KK \cap \dom(L)$ belongs to $\KK_*$ if, and only if,
\begin{equation}
	\bs{b}_i^\top \nabla L(\bs{\psi}) \ \begin{cases}
		=   \ 0 & \text{for all} \ i \in \{1,\ldots,m\} \setminus A(\bs{\psi}) , \\
		\le \ 0 & \text{for all} \ i \in A(\bs{\psi}) .
	\end{cases}
	\label{eq: KKstar}
\end{equation}

\noindent
\textbf{(b)} For any given set $A \subseteq \{1,\ldots,q\}$, a vector $\bs{\psi} \in \VV(A) \cap \dom(L)$ belongs to $\VV_*(A)$ if, and only if,
\begin{equation}
	\bs{b}_i^\top \nabla L(\bs{\psi}) \ = \ 0
		\quad\mbox{for all } \ i \in \{1,\ldots,m\} \setminus A .
	\label{eq: VVA}
\end{equation}
\end{Theorem}

The characterizations in this theorem entail that any vector $\bs{\psi} \in \KK_*$ belongs to $\VV_*(A(\bs{\psi}))$. The active set algorithm performs one of the following two procedures alternately:

\paragraph{Basic procedure 1: Replacing a feasible point with a ``conditionally'' optimal one.}
Let $\bs{\psi}$ be an arbitrary vector in $\KK \cap \dom(L)$. Our goal is to find a vector $\bs{\psi}_{\rm new}$ such that
\begin{equation}
	L(\bs{\psi}_{\rm new}) \ \ge \ L(\bs{\psi})
	\quad\mbox{and}\quad
	\bs{\psi}_{\rm new} \ \in \ \KK \cap \VV_*(A(\bs{\psi}_{\rm new})) .
	\label{eq: local optimality}
\end{equation}
To this end, set $A := A(\bs{\psi})$ and define the candidate vector $\bs{\psi}_{\rm cand} := \til{\bs{\psi}}(A)$. By construction, $L(\bs{\psi}_{\rm cand}) \ge L(\bs{\psi})$. If $L(\bs{\psi}_{\rm cand}) = L(\bs{\psi})$, we set $\bs{\psi}_{\rm new} := \bs{\psi}$. If $L(\bs{\psi}_{\rm cand}) > L(\bs{\psi})$ and $\bs{\psi}_{\rm cand} \in \KK$, we set $\bs{\psi}_{\rm new} := \bs{\psi}_{\rm cand}$. Here \eqref{eq: local optimality} is satisfied, because $A(\bs{\psi}_{\rm new}) \supseteq A(\bs{\psi})$, so that $\VV(A(\bs{\psi}_{\rm new})) \subseteq \VV(A)$. Finally, if $L(\bs{\psi}_{\rm cand}) > L(\bs{\psi})$ but $\bs{\psi}_{\rm cand} \not\in \KK$, let
\begin{eqnarray}
	t = t(\bs{\psi}, \bs{\psi}_{\rm cand})
	& := & \max \bigl\{ t \in (0,1) : (1 - t) \bs{\psi} + t \bs{\psi}_{\rm cand} \in \KK \bigr\}
	\label{eq:def of t} \\
	& = & \min \Bigl\{ \frac{c_i - \bs{v}_i^\top \bs{\psi}}
	                      {\bs{v}_i^\top \bs{\psi}_{\rm cand} - \bs{v}_i^\top \bs{\psi}} :
			1 \le i \le q, \bs{v}_i^\top \bs{\psi}_{\rm cand} > c_i \Bigr\} .
	\nonumber
\end{eqnarray}
Then we replace $\bs{\psi}$ with $(1 - t)\bs{\psi} + t \bs{\psi}_{\rm cand}$. Note that $L(\bs{\psi})$ does not decrease in this step, due to concavity of $L$. Moreover, the set $A(\bs{\psi})$ increases strictly. Hence, repeating the preceding manipulations at most $q$ times yields finally a vector $\bs{\psi}_{\rm new}$ satisfying \eqref{eq: local optimality}, because $\VV(\{1,\ldots,q\})$ is clearly a subset of $\KK$. With the new vector $\bs{\psi}_{\rm new}$ we perform the second basic procedure.

\paragraph{Basic procedure 2: Altering the set of active constraints.}
Let $\bs{\psi} \in \KK \cap \dom(L) \cap \VV_*(A)$ with $A = A(\bs{\psi})$. It follows from Theorem~\ref{thm: KKstar and VVA} that $\bs{\psi}$ belongs to $\KK_*$ if, and only if,
$$
	\bs{b}_a^\top \nabla L(\bs{\psi}) \ \le \ 0
	\quad\mbox{for all } a \in A .
$$
Now suppose that the latter condition is violated, and let $a_o = a_o(\bs{\psi})$ be an index in $A$ such that $\bs{b}_{a_o}^\top \nabla L(\bs{\psi})$ is maximal. Then $\bs{\psi} + t \bs{b}_{a_o} \in \KK$ and $A(\bs{\psi} + t \bs{b}_{a_o}) = A \setminus \{a_o\}$ for arbitrary $t > 0$, while $L(\bs{\psi} + t \bs{b}_{a_o}) > L(\bs{\psi})$ for sufficiently small $t > 0$. Thus we consider the vector $\bs{\psi}_{\rm cand} := \til{\bs{\psi}}(A \setminus \{a_o\})$, which satisfies necessarily the inequality $L(\bs{\psi}_{\rm cand}) > L(\bs{\psi})$. It may fail to be in $\KK$, but it satisfies the inequality
$$
	\bs{v}_{a_o}^\top \bs{\psi}_{\rm cand} \ > \ c_{a_o} .
$$
For $\bs{\psi}_{\rm cand} - \bs{\psi}$ may be written as $\lambda_{a_o} \bs{b}_{a_o} + \sum_{i \not\in A} \lambda_i \bs{b}_i$ with real coefficients $\lambda_1,\ldots,\lambda_m$, and
$$
	0 \ < \ (\bs{\psi}_{\rm cand} - \bs{\psi})^\top \nabla L (\bs{\psi})
	\ = \ \lambda_{a_o} \bs{b}_{a_o}^\top \nabla L(\bs{\psi})
$$
according to \eqref{eq: VVA}. Hence $0 < \lambda_{a_o} = \bs{v}_{a_o}^\top(\bs{\psi}_{\rm cand} - \bs{\psi}) = \bs{v}_{a_o}^\top\bs{\psi}_{\rm cand} - c_{a_o}$. If $\bs{\psi}_{\rm cand} \in \KK$, we repeat this procedure with $\bs{\psi}_{\rm cand}$ in place of $\bs{\psi}$. Otherwise, we replace $\bs{\psi}$ with $(1 - t) \bs{\psi} + t \bs{\psi}_{\rm cand}$, where $t = t(\bs{\psi}, \bs{\psi}_{\rm cand}) > 0$ is defined in \eqref{eq:def of t}, which results in a strictly larger value of $L(\bs{\psi})$. Then we perform the first basic procedure.

\paragraph{The complete algorithm and its validity.}
Often one knows a vector $\bs{\psi}_o \in \KK \cap \dom(L)$ in advance. Then the active set algorithm can be started with the first basic procedure and proceeds as indicated in Table~\ref{t: Active Set 1}. In other applications it is sometimes obvious that $\VV(\{1,\ldots,q\})$, which is clearly a subset of $\KK$, contains a point in $\dom(L)$. In that case the input vector $\bs{\psi}_o$ is superfluous, and the first twelve lines in Table~\ref{t: Active Set 1} may be simplified as indicated in Table~\ref{t: Active Set 2}. The latter approach with starting point $\bs{\psi}_o = \til{\bs{\psi}}(\{1,\ldots,q\})$ may be numerically unstable, presumably when this starting point is very far from the optimum. In the special settings of concave least squares regression or log-concave density estimation, a third variant turned out to be very reliable: We start with $A = \emptyset$ and $\bs{\psi}_o = \til{\bs{\psi}}(A)$. As long as $\bs{\psi}_o \not\in \KK$, we replace $A$ with the larger set $A(\bs{\psi}_o)$ and recompute $\bs{\psi}_o = \til{\bs{\psi}}(A)$; see Table~\ref{t: Active Set 3}.

In Table~\ref{t: Active Set 1}, the lines marked with (*) and (**) correspond to the end of the first basic procedure. At this stage, $\bs{\psi}$ is a vector in $\KK \cap \dom(L) \cap \VV_*(A(\bs{\psi}))$. Moreover, whenever the point (**) is reached, the value $L(\bs{\psi})$ is strictly larger than previously and equal to the maximum of $L$ over the set $\VV(A)$. Since there are only finitely many different sets $A \subseteq \{1,\ldots,q\}$, the algorithm terminates after finitely many steps, and the resulting $\bs{\psi}$ belongs to $\KK$ by virtue of Theorem~\ref{thm: KKstar and VVA}.

When implementing these algorithms one has to be aware of numerical inaccuracies and errors, in particular, if the algorithm $\til{\bs{\psi}}(\cdot)$ yields only approximations of vectors in $\VV_*(\cdot)$. In our specific applications we avoided endless loops by replacing the conditions ``$\bs{b}_a^\top \nabla L(\bs{\psi}) < 0$'' and ``$\bs{v}_i^\top \bs{\psi} > c_i$'' with ``$\bs{b}_a^\top \nabla L(\bs{\psi}) < - \eps$'' and ``$\bs{v}_i^\top \bs{\psi} > c_i + \eps$'', respectively, for some small constant $\eps > 0$.

\begin{table}[h]
\centerline{\bf\begin{tabular}{|l|} \hline
\ruck{0}	Algorithm $\bs{\psi} \leftarrow \mbox{ActiveSet1}(L,\til{\bs{\psi}}^{\strut}(\cdot),\bs{\psi}_o)$\\
\ruck{0}	$\bs{\psi} \leftarrow \bs{\psi}_o$\\
\ruck{0}	$A \leftarrow A(\bs{\psi})$\\
\ruck{0}	$\bs{\psi}_{\rm cand} \leftarrow \til{\bs{\psi}}(A)$\\
\ruck{0}	while $\bs{\psi}_{\rm cand} \not\in \KK$ do\\
\ruck{1}		$\bs{\psi} \leftarrow (1 - t(\bs{\psi},\bs{\psi}_{\rm cand})) \bs{\psi}
					+ t(\bs{\psi},\bs{\psi}_{\rm cand}) \bs{\psi}_{\rm cand}$\\
\ruck{1}		$A \leftarrow A(\bs{\psi})$\\
\ruck{1}		$\bs{\psi}_{\rm cand} \leftarrow \til{\bs{\psi}}(A)$\\
\ruck{0}	end while\\
\ruck{0}	$\bs{\psi} \leftarrow \bs{\psi}_{\rm cand}$\\
\ruck{0}	$A \leftarrow A(\bs{\psi})$	\quad	(*)\\
\ruck{0}	while $\max_{a \in A} \bs{b}_a^\top \nabla L(\bs{\psi}) > 0$ do\\
\ruck{1}		$a \leftarrow \min \left( \mathop{\rm argmax}_{a \in A} \bs{b}_a^\top \nabla L(\bs{\psi}) \right)$\\
\ruck{1}		$A \leftarrow A \setminus \{a\}$\\
\ruck{1}		$\bs{\psi}_{\rm cand} \leftarrow \til{\bs{\psi}}(A)$\\
\ruck{1}		while $\bs{\psi}_{\rm cand} \not\in \KK$ do\\
\ruck{2}			$\bs{\psi} \leftarrow (1 - t(\bs{\psi},\bs{\psi}_{\rm cand})) \bs{\psi}
						+ t(\bs{\psi},\bs{\psi}_{\rm cand}) \bs{\psi}_{\rm cand}$\\
\ruck{2}			$A \leftarrow A(\bs{\psi})$\\
\ruck{2}			$\bs{\psi}_{\rm cand} \leftarrow \til{\bs{\psi}}(A)$\\
\ruck{1}		end while\\
\ruck{1}		$\bs{\psi} \leftarrow \bs{\psi}_{\rm cand}$\\
\ruck{1}		$A \leftarrow A(\bs{\psi})$	\quad	(**)\\
\ruck{0}	end while.\\\hline
\end{tabular}}
\caption{Pseudo-code of an active set algorithm.}
\label{t: Active Set 1}
\end{table}

\begin{table}[h]
\centerline{\bf\begin{tabular}{|l|} \hline
\ruck{0}	Algorithm $\bs{\psi} \leftarrow \mbox{ActiveSet2}(L,\til{\bs{\psi}}^{\strut}(\cdot))$\\
\ruck{0}	$\bs{\psi} \leftarrow \til{\bs{\psi}}(\{1,\ldots,q\})$\\
\ruck{0}	$A \leftarrow \{1,\ldots,q\}$\\
\ruck{0}	while $\max_{a \in A} \bs{b}_a^\top \nabla L(\bs{\psi}) > 0$ do\\
\ruck{1}		\ldots\\
\ruck{0}	end while.\\\hline
\end{tabular}}
\caption{Pseudo-code of first modified active set algorithm.}
\label{t: Active Set 2}
\end{table}

\begin{table}[h]
\centerline{\bf\begin{tabular}{|l|} \hline
\ruck{0}	Algorithm $\bs{\psi} \leftarrow \mbox{ActiveSet3}(L,\til{\bs{\psi}}^{\strut}(\cdot))$\\
\ruck{0}	$\bs{\psi} \leftarrow \til{\bs{\psi}}(\emptyset)$\\
\ruck{0}	while $\bs{\psi} \not\in \KK$ do\\
\ruck{1}		$A \leftarrow A(\bs{\psi})$\\
\ruck{1}		$\bs{\psi} \leftarrow \til{\bs{\psi}}(A)$\\
\ruck{0}	end while\\
\ruck{0}	$A \leftarrow A(\bs{\psi})$\\
\ruck{0}	while $\max_{a \in A} \bs{b}_a^\top \nabla L(\bs{\psi}) > 0$ do\\
\ruck{1}		\ldots\\
\ruck{0}	end while.\\\hline
\end{tabular}}
\caption{Pseudo-code of second modified active set algorithm.}
\label{t: Active Set 3}
\end{table}

\paragraph{Possible extension I.}
The assumption of linearly independent vectors $\bs{v}_1, \ldots, \bs{v}_q$ has been made for convenience and could be relaxed of course. In particular, one can extend the previous considerations easily to the situation where $\KK$ consists of all vectors $\bs{\psi} \in \R^m$ such that
$$
	c_{i,1} \ \le \ \bs{v}_i^\top \bs{\psi} \ \le \ c_{i,2}
$$
for $1 \le i \le q$ with numbers $- \infty \le c_{i,1} < c_{i,2} < \infty$.

\paragraph{Possible extension II.}
Again we drop assumption~\eqref{ass: linear independence} but assume that $c_1 = \cdots = c_q = 0$, so that $\KK$ is a closed convex cone. Suppose further that we know a finite set $\mathcal{E}$ of generators of $\KK$, i.e.\ every vector $\bs{\psi} \in \KK$ may be written as
$$
	\bs{\psi} \ = \ \sum_{\bs{e} \in \mathcal{E}} \lambda_{\bs{e}} \bs{e}
$$
with numbers $\lambda_{\bs{e}} \ge 0$. In that case, a point $\bs{\psi} \in \KK \cap \mathrm{dom}(L)$ belongs to $\KK_*$ if, and only if,
\begin{equation}
	\nabla L(\bs{\psi})^\top \bs{\psi} \ = \ 0
	\quad\text{and}\quad
	\max_{\bs{e} \in \mathcal{E}} \, \nabla L(\bs{\psi})^\top \bs{e} \ \le \ 0 .
	\label{eq: KKstar2}
\end{equation}
Now we can modify our basic procedure~2 as follows: Let $\bs{\psi} \in \KK \cap \mathrm{dom}(L) \cap \VV(A)$ with $A := A(\bs{\psi})$. If \eqref{eq: KKstar2} is violated, let $\bs{e}(\bs{\psi}) \in \mathcal{E}$ such that $\nabla L(\bs{\psi})^\top \bs{e}(\bs{\psi}) > 0$. Further let $s(\bs{\psi}), t(\bs{\psi}) > 0$ such that $\bs{\psi}_{\rm new} := s(\bs{\psi}) \bs{\psi} + t(\bs{\psi}) \bs{e}(\bs{\psi}) \in \KK$ satisfies $L(\bs{\psi}_{\rm new}) > L(\bs{\psi})$. Then we replace $\bs{\psi}$ with $\bs{\psi}_{\rm new}$ and perform the first basic procedure.

\subsection{The special case of fitting log-concave densities}

Going back to our original problem, note that $\psi \in \GG$ lies within $\GGconc$ if, and only if, the corresponding vector $\bs{\psi}$ satisfies
\begin{equation}
	\frac{\psi_{j+1} - \psi_j}{\delta_j} - \frac{\psi_j - \psi_{j-1}}{\delta_{j-1}}
		= \bs{v}_j^\top \bs{\psi}
	\ \le \ 0	\quad\mbox{for } j = 2,\ldots,m-1 ,
	\label{eq: concavity of psi}
\end{equation}
where $\bs{v}_j = (v_{i,j})_{i=1}^m$ has exactly three nonzero components:
$$
	v_{j-1,j} \ := \ 1/\delta_{j-1} ,	\quad
	v_{j,j} \ := \ - (\delta_{j-1} + \delta_j)/(\delta_{j-1} \delta_j) ,	\quad
	v_{j+1,j} \ := \ 1/\delta_j .
$$
Note that we changed the notation slightly by numbering the $m - 2$ constraint vectors from $2$ to $m-1$. This is convenient, because then $\bs{v}_j^\top \bs{\psi} \ne 0$ is equivalent to the corresponding function $\psi \in \GG$ changing slope at $x_j$. Suitable basis vectors $\bs{b}_i$ are given, for instance, by $\bs{b}_1 := (1)_{i=1}^m$, $\bs{b}_m := (x_i)_{i=1}^m$ and
$$
	\bs{b}_j \ = \ \bigl( \min(x_i - x_j, 0) \bigr)_{i=1}^m,	\quad	2 \le j < m .
$$

For this particular problem it is convenient to rephrase the active set method in terms of {\sl inactive} constraints, i.e.\ true {\sl knots} of functions in $\GG$. Throughout let $I = \{i(1),\ldots,i(k)\}$ be a subset of $\{1,2,\ldots,m\}$ with $k \ge 2$ elements $1 = i(1) < \cdots < i(k) = m$, and let $\GG(I)$ be the set of all functions $\psi \in \GG$ which are linear on all intervals $[x_{i(s)}, x_{i(s+1)}]$, $1 \le s < k$. This set corresponds to $\VV(A)$ with $A := \{1,\ldots,m\} \setminus I$. A function $\psi \in \GG(I)$ is uniquely determined by the vector $\bigl( \psi(x_{i(s)}) \bigr)_{s=1}^k$, and one may write
$$
	L(\psi) \ = \ \sum_{s=1}^k p_s(I) \psi(x_{i(s)})
		- \sum_{s=1}^{k-1} (x_{i(s+1)} - x_{i(s)}) J \bigl( \psi(x_{i(s)}), \psi(x_{i(s+1)}) \bigr)
$$
with suitable probability weights $p_1(I), \ldots, p_k(I) > 0$. Precisely, writing
$$
	\psi(x)
	\ = \ \frac{x_{i(s+1)} - x}{x_{i(s+1)} - x_{i(s)}} \, \psi(x_{i(s)})
		+ \frac{x - x_{i(s)}}{x_{i(s+1)} - x_{i(s)}} \, \psi(x_{i(s+1)})
$$
for $1 \le s < k$ and $x_{i(s)} \le x \le x_{i(s+1)}$ yields the explicit formulae
\begin{eqnarray*}
	p_1(I)
		& = & \sum_{i=1}^{i(2)-1} \frac{x_{i(2)} - x_i}{x_{i(2)} - x_1} \, p_i^{} , \\
	p_s(I)
		& = & \sum_{i=i(s-1)+1}^{i(s+1)-1}
			\min \Bigl( \frac{x_i - x_{i(s-1)}}{x_{i(s)} - x_{i(s-1)}},
			          \frac{x_{i(s+1)} - x_i}{x_{i(s+1)} - x_{i(s)}} \Bigr) \,
			p_i^{}
			\quad\mbox{for } 2 \le s < k ,\\
	p_k(I)
		& = & \sum_{i=i(k-1)+1}^m \frac{x_i - x_{i(k-1)}}{x_m - x_{i(k-1)}} \, p_i^{} .
\end{eqnarray*}
Consequently, the computation of $\til{\psi}$ or $\til{\psi}^{(I)} := \mathop{\rm argmax}_{\psi \in \GG(I)} L(\psi)$ are optimization problems of the same type.

Since the vectors $\bs{b}_2,\ldots,\bs{b}_m$ correspond to the functions $\Delta_2, \ldots, \Delta_m$ in $\GG$ with
\begin{equation}
	\Delta_j(x) \ := \ \min(x - x_j, 0) ,
	\label{eq: def Del_j}
\end{equation}
checking the inequality $\bs{b}_a^\top \nabla L(\bs{\psi}) \le 0$ for $a \in A$ amounts to checking whether the directional derivative
\begin{equation}
	H_j(\psi) \ := \ \sum_{i=1}^m p_i \Delta_j(x_i) - \int_{x_1}^{x_m} \Delta_j(x) \exp \psi(x) \, dx
	\label{eq: def H_j}
\end{equation}
is nonpositive for all $j \in \{1,\ldots,m\} \setminus I$. If $\psi = \psi^{(I)}$ and $j \not\in I$, the inequality $H_j(\psi) > 0$ means that $L(\psi)$ could be increased strictly by allowing an additional knot at $x_j$.

\begin{Example}
\label{ex: Example 1}
Figure~\ref{fig: Example_F} shows the empirical distribution function of $n = 25$ simulated random variables from a Gumbel distribution, while the smooth distribution function is the estimator $\hat{F}(r) := \int_{-\infty}^r \exp \hat{\psi}(x) \, dx$. Figure~\ref{fig: Example_phi} illustrates the computation of the log-density $\hat{\psi}$ itself. Each picture shows the current function $\psi$ together with the new candidate function $\psi_{\rm cand}$. We followed the algorithm in Table~\ref{t: Active Set 2}, so the first (upper left) picture shows the starting point, a linear function $\psi$ on $[x_1, x_{25}]$, together with $\psi_{\rm cand}$ having an additional knot in $(x_1,x_{25})$. Since $\psi_{\rm cand}$ is concave, it becomes the new function $\psi$ shown in the second (upper right) plot. In the third (lower left) plot one sees the situation where adding another knot resulted in a non-concave function $\psi_{\rm cand}$. So the current function $\psi$ was replaced with a convex combination of $\psi$ and $\psi_{\rm cand}$. The latter new function $\psi$ and the almost identical final fit $\hat{\psi}$ are depicted in the fourth (lower right) plot.

\begin{figure}[h]
\centerline{\includegraphics[width=10cm]{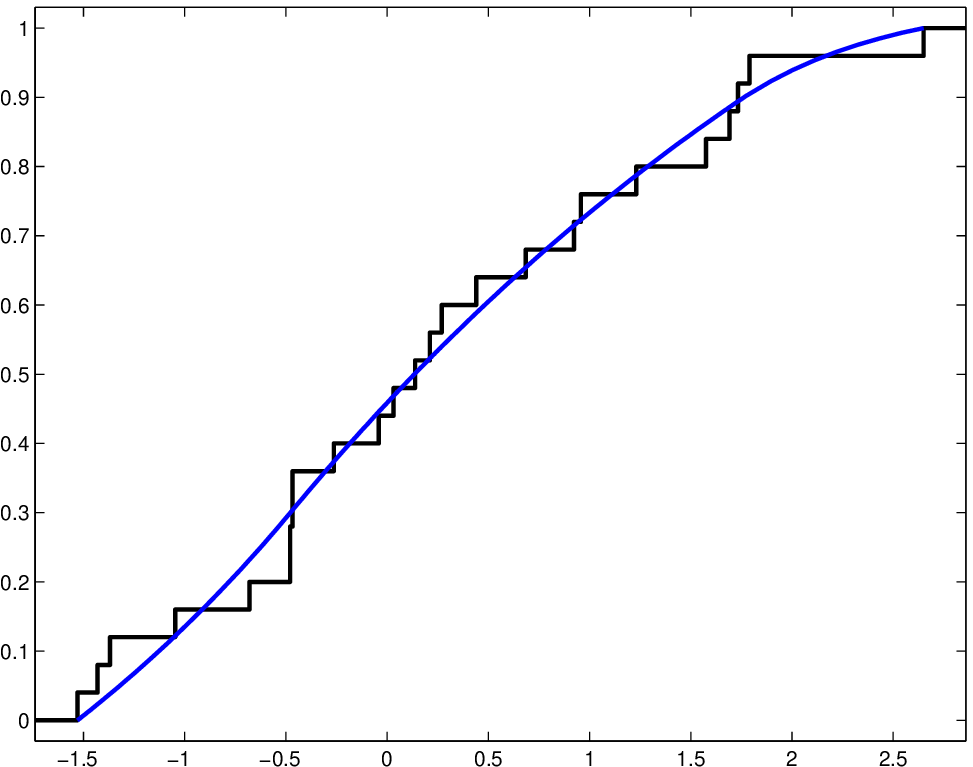}}
\caption{Estimated distribution functions for $n = 25$ data points.}
\label{fig: Example_F}
\end{figure}

\begin{figure}[h]
\includegraphics[width=7.4cm]{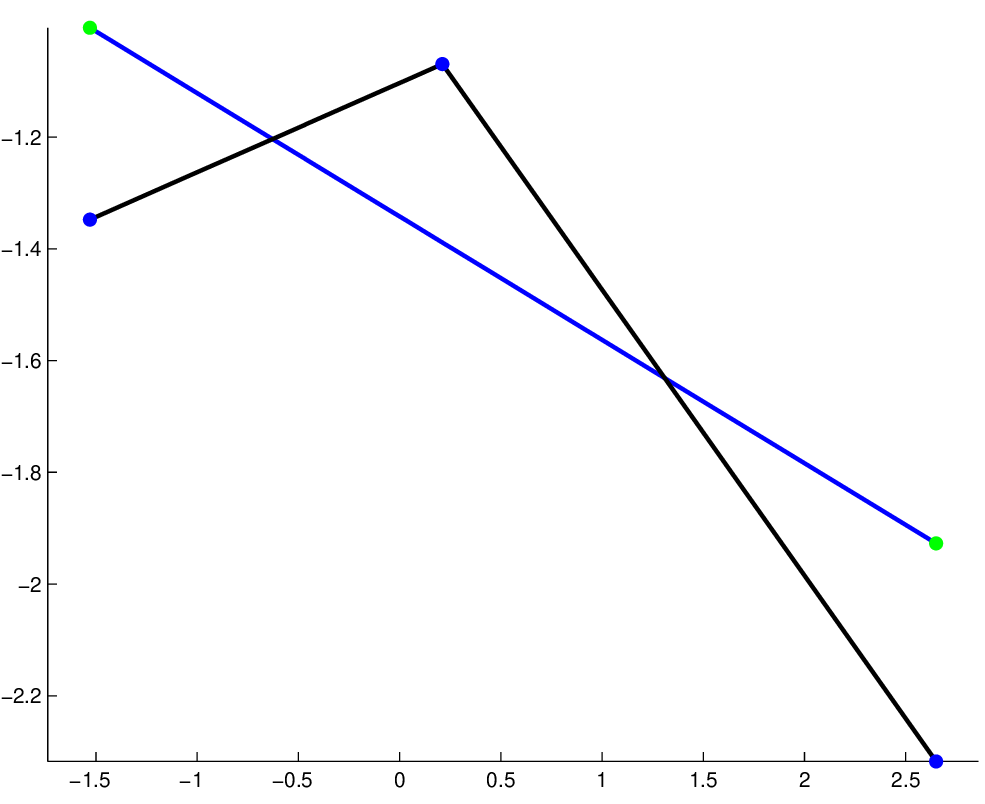}
\hfill
\includegraphics[width=7.4cm]{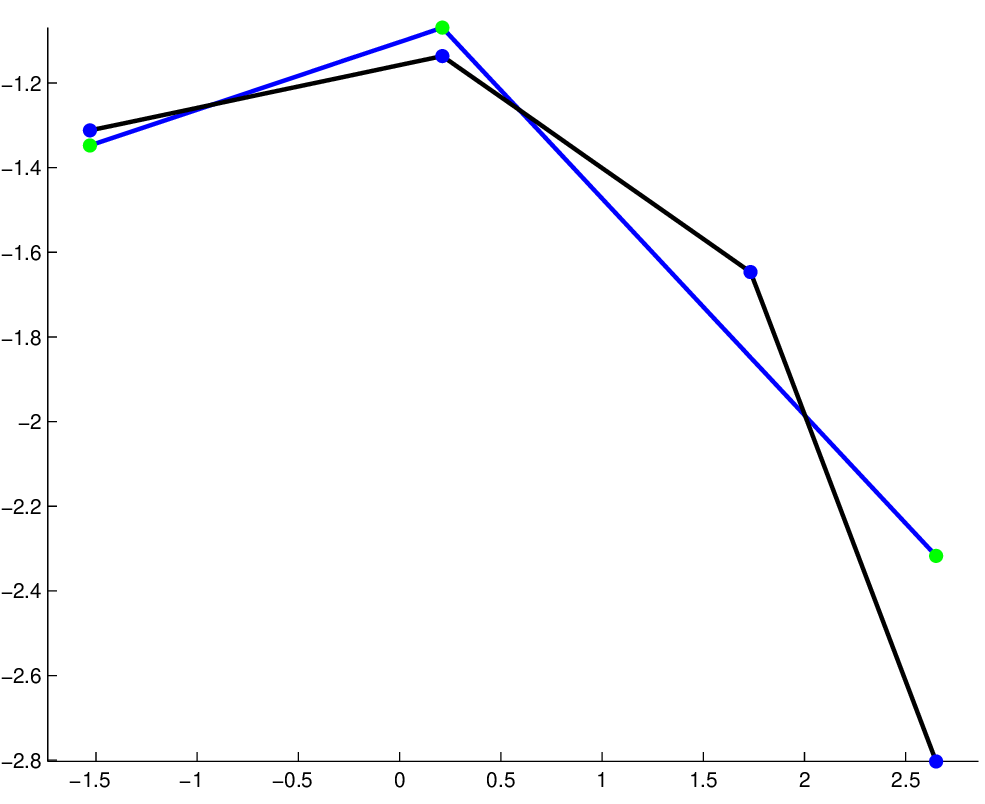}

\includegraphics[width=7.4cm]{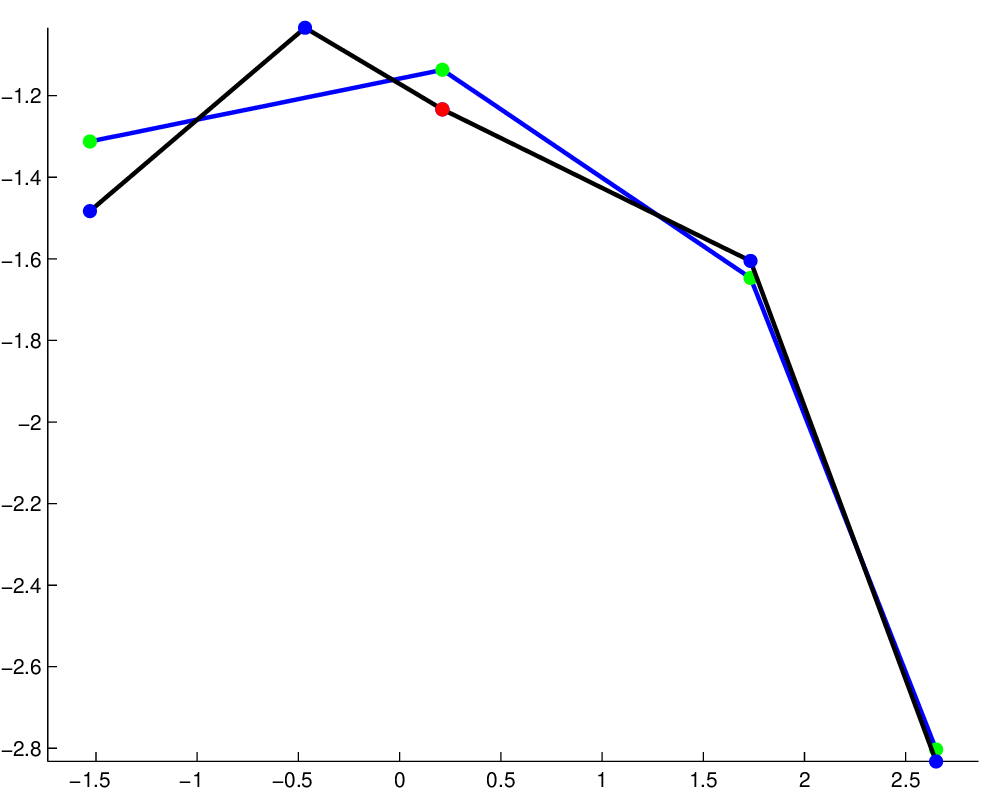}
\hfill
\includegraphics[width=7.4cm]{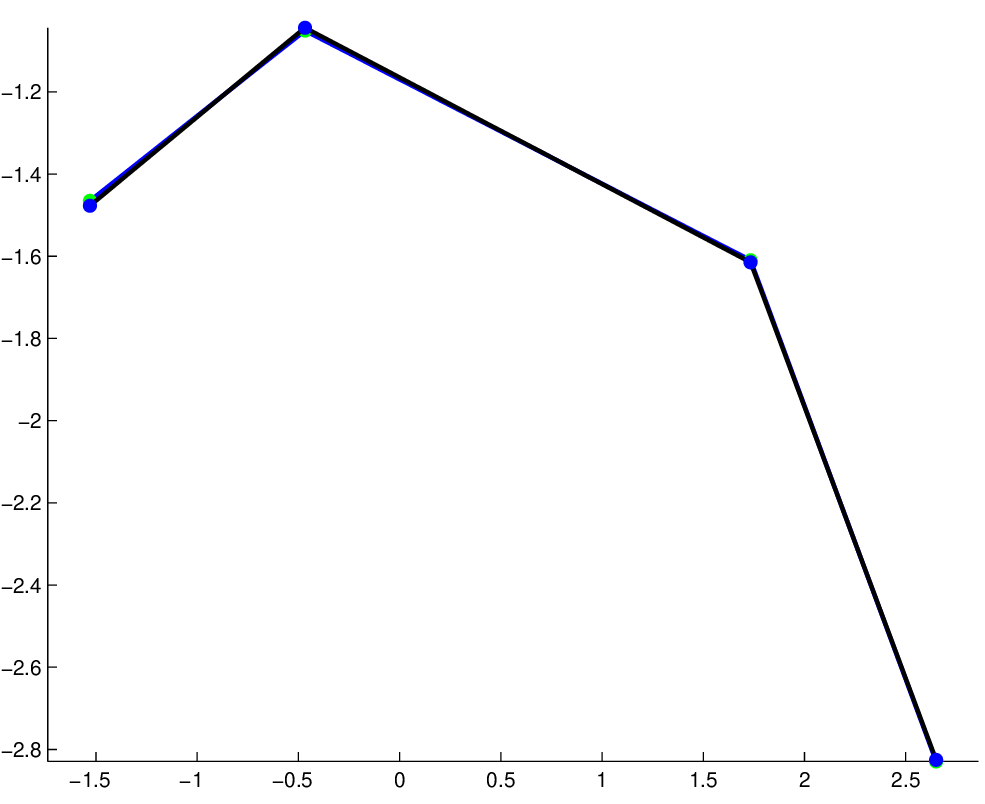}

\caption{Estimating the log-density for $n = 25$ data points.}
\label{fig: Example_phi}
\end{figure}
\end{Example}

\section{Censored or binned data}
\label{Censored Data}

In the current and the next section we consider independent random variables $X_1$, $X_2$, \ldots, $X_n$ with unknown distribution $P$ on $(0,\infty]$ having sub-probability density $f = \exp\phi$ on $(0,\infty)$, where $\phi$ is concave and upper semicontinuous. In many applications the observations $X_i$ are not completely available. For instance, let the $X_i$ be event times for $n$ individuals in a biomedical study, where $X_i = \infty$ means that the event in question does not happen at all. If the study ends at time $c_i > 0$ from the $i$-th unit's viewpoint, whereas $X_i > c_i$, then we have a ``right-censored'' observation and know only that $X_i$ is contained in the interval $\til{X}_i = (c_i, \infty]$. In other settings one has purely ``interval-censored'' data: For the $i$-th observation one knows only which of given intervals $(0,t_{i,1}], (t_{i,1}, t_{i,2}], \ldots, (t_{i,m(i)},\infty]$ contains $X_i$, where $0 < t_{i,1} < \cdots < t_{i,m(i)} < \infty$. If these candidate intervals are the same for all observations, one speaks of binned data. A related situation are rounded observations, e.g.\ when we observe $\lceil X_i \rceil$ rather than $X_i$.

In all these settings we observe independent random intervals $\til{X}_1$, $\til{X}_2$, \ldots, $\til{X}_n$. More precisely, we assume that either $\til{X}_i = (L_i, R_i] \ni X_i$ with $0 \le L_i < R_i \le \infty$, or $\til{X}_i$ consists only of the one point $L_i := R_i := X_i \in (0,\infty)$. The normalized log-likelihood for this model reads
\begin{eqnarray}
\label{eq: log-likelihood censored}
	\bar{\ell}(\phi)
	& := & n_{}^{-1} \sum_{i=1}^n 
		\biggl( 1\{L_i = R_i\} \phi(X_i) \\
	&& \qquad\qquad\qquad + \ 1\{L_i < R_i\}
		\log \Bigl( \int_{L_i}^{R_i} \exp \phi(x) \, dx + 1\{R_i = \infty\} p_\infty \Bigr) \biggr) ,
	\nonumber
\end{eqnarray}
where
$$
	p_\infty \ := \ 1 - \int_0^\infty \exp\phi(x) \, dx
	\ \in \ [0,1] .
$$

\section{An EM algorithm}
\label{EM}

Maximizing the log-likelihood function $\bar{\ell}(\phi)$ for censored data is a non-trivial task and will be treated in detail elsewhere. Here we only indicate how this can be achieved in principle, assuming for simplicity that $P(\{\infty\}) = 0$, i.e.\ $\int_0^\infty \exp\phi(x) \, dx = 1$ and $p_\infty = 0$. In this case, the log-likelihood simplifies to
$$
	\bar{\ell}(\phi)
	\ = \ n_{}^{-1} \sum_{i=1}^n 
		\biggl( 1\{L_i = R_i\} \phi(X_i)
			+ 1\{L_i < R_i\}
				\log \Bigl( \int_{L_i}^{R_i} \exp \phi(x) \, dx \Bigr) \biggr) .
$$
Again one may get rid of the constraint $\int_0^\infty \exp\phi(x) \, dx = 1$ by considering
\begin{equation}
\label{eq: log-likelihood censored simple}
	\bar{L}(\phi) \ := \ \bar{\ell}(\phi) - \int_0^\infty \exp\phi(x) \, dx
\end{equation}
for arbitrary concave and upper semicontinuous functions $\phi : (0,\infty) \to [-\infty,\infty)$.

A major problem is that $\bar\ell(\phi)$ is not linear but convex in $\phi$. Namely, for $v : (0,\infty) \to \R$ and $0 \le L < R \le \infty$,
\begin{equation}
\label{eq: dir deriv}
	\frac{d^a}{dt^a} \Big\vert_{t=0}^{}
		\log \Bigl( \int_L^R \exp(\psi(x) + t v(x)) \, dx \Bigr)
	\ = \ \begin{cases}
			\Ex_\phi \bigl( v(X) \,\big|\, X \in (L,R] \bigr) & \text{if} \ a = 1 , \\
			\Var_\phi \bigl( v(X) \,\big|\, X \in (L,R] \bigr) & \text{if} \ a = 2 .
		\end{cases}
\end{equation}
Thus we propose to maximize $\bar{\ell}(\phi)$ iteratively as follows: Starting from a function $\phi$ with $\bar{L}(\phi) > - \infty$, we replace the target function $\bar{L}(\phi_{\rm new})$ with
$$
	\til{L}(\phi_{\rm new} \,|\, \phi)
	\ := \ \frac{d}{dt} \Big\vert_{t=0}^{} \bar{\ell} \bigl( \phi + t (\phi_{\rm new} - \phi) \bigr)
		- \int_0^\infty \exp\phi_{\rm new}(x) \, dx .
$$
By means of (\ref{eq: dir deriv}), this may be written as
\begin{equation}
\label{eq: surrogate}
	\til{L}(\phi_{\rm new} \,|\, \phi)
	\ = \ \mathrm{const}(\phi) + \int \phi_{\rm new}(x) \, P(dx \,|\, \phi)
		- \int_0^\infty \exp\phi_{\rm new}(x) \, dx ,
\end{equation}
where
$$
	P(\cdot \,|\, \phi)
	\ := \ n_{}^{-1} \sum_{i=1}^n \biggl( 1\{L_i = R_i\} \delta_{X_i}^{}
		+ 1\{L_i < R_i\} \mathcal{L}_\phi \bigl( X \,\big|\, X \in (L_i,R_i] \bigr) \biggr) ,
$$
a probability measure depending on the data and on $\phi$. In other words, for any Borel subset $B$ of $(0,\infty)$,
$$
	P(B \,|\, \phi)
	\ := \ n_{}^{-1} \sum_{i=1}^n \biggl( 1\{L_i = R_i \in B\}
		+ 1\{L_i < R_i\}
			\frac{\int_{B \cap (L_i,R_i)} \exp\phi(x) \, dx}{\int_{(L_i,R_i)} \exp\phi(x) \, dx} \biggr) .
$$
Note also that $\til{L}(\phi_{\rm new} \,|\, \phi)$ equals the conditional expectation of the complete-data log-likelihood $L(\phi_{\rm new})$, given the available data and assuming the current $\phi$ to be the true log-density:
$$
	\til{L}(\phi_{\rm new} \,|\, \phi)
	\ = \ \Ex_\phi \bigl( L(\phi_{\rm new}) \,\big|\, X_i \in \til{X}_i \ \text{for} \ 1 \le i \le n \bigr) ,
$$
where the $\til{X}_i$ are treated temporarily as fixed.
 
After approximating the probability measure $P(\cdot \,|\, \phi)$ by a discrete distribution with finite support, one can maximize $\til{L}(\phi_{\rm new} \,|\, \phi)$ over all concave functions $\phi_{\rm new}$ with the active-set algorithm presented in Section~\ref{Active Set}. Then we replace $\phi$ with $\phi_{\rm new}$ and repeat this procedure until the change of $\phi$ becomes negligable.

\section{Auxiliary results and proofs}
\label{Proofs}

\paragraph{Explicit formulae for $J$ and some of its partial derivatives.}
Recall the auxiliary function $J(r,s) := \int_0^1 \exp((1 - t)r + ts) \, dt$. One may write explicitly
$$
	J(r,s) = J(s,r) \ = \ \begin{cases}
		\bigl( \exp(r) - \exp(s) \bigr) \big/ (r - s) & \mbox{if } r \ne s , \\
		\exp(r)                                       & \mbox{if } r = s ,
	\end{cases}
$$
or utilize the fact that $J(r,s) = \exp(r) J(0, s-r)$ with $J(0,0) = 1$ and
$$
	J(0, y) \ = \ (\exp(y) - 1) / y
	\ = \ \sum_{k=0}^\infty \frac{y^k}{(k+1)!} .
$$

To compute the partial derivatives $J_{ab}(r,s)$ of $J(r,s)$, one may utilize the facts that $J_{ab}(r,s) = J_{ba}(s,r) = \exp(r) J_{ab}(0, s-r)$. Moreover, elementary calculations reveal that
\begin{eqnarray*}
	J_{10}(0,y)
	& = & \bigl( \exp(y) - 1 - y \bigr) \big/ y^2
		\ = \ \sum_{k=0}^\infty \frac{y^k}{(k+2)!} , \\
	J_{20}(0,y)
	& = & 2 \bigl( \exp(y) - 1 - y - y^2/2 \bigr) \big/ y^3
		\ = \ \sum_{k=0}^\infty \frac{2 y^k}{(k+3)!} , \\
	J_{11}(0,y)
	& = & \bigl( y (\exp(y) + 1) - 2 (\exp(y) - 1) \bigr) \big/ y^3
		\ = \ \sum_{k=0}^\infty \frac{(k+1) y^k}{(k+3)!} .
\end{eqnarray*}
The Taylor series may be deduced as follows:
\begin{eqnarray*}
	J_{ab}(0,y)
	& = & \int_0^1 (1 - t)^a t^b e^{ty} \, dt \\
	& = & \sum_{k=0}^\infty \frac{y^k}{k!} \int_0^1 (1 - t)^a t^{b+k} \, dt \\
	& = & \sum_{k=0}^\infty \frac{y^k}{k!} \frac{a! (b+k)!}{(k + a + b + 1)!} \\
	& = & \sum_{k=0}^\infty \frac{a! (b+k)! \, y^k}{k! (k + a + b + 1)!} ,
\end{eqnarray*}
according to the general formula $\int_0^1 (1 - t)^k t^\ell \, dt = k! \ell! / (k+\ell + 1)!$ for integers $k,\ell \ge 0$.

Numerical experiments revealed that a fourth degree Taylor approximation for $J_{ab}(0,y)$ is advisable and works very well if
$$
	|y| \ \le \ \begin{cases}
		0.005 & (a = b = 0) , \\
		0.01  & (a + b = 1) , \\
		0.02  & (a + b = 2) .
	\end{cases}
$$

\paragraph{Explicit formulae for the gradient and hessian matrix of $L$.}
At $\bs{\psi} \in \R^m$ these are given by
\begin{eqnarray*}
	\frac{\partial}{\partial \psi_k} L(\bs{\psi})
	& = & p_k^{} - \begin{cases}
			\delta_1 J_{10}(\psi_1, \psi_2)
				& \mbox{if } k = 1 , \\
			\delta_{k-1} J_{01}(\psi_{k-1},\psi_k) + \delta_k J_{10}(\psi_k,\psi_{k+1})
				& \mbox{if } 2 \le k < m , \\
			\delta_{m-1} J_{01}(\psi_{m-1},\psi_m)
				& \mbox{if } k = m ,
		\end{cases} \\
	- \, \frac{\partial^2}{\partial \psi_j \partial \psi_k} L(\bs{\psi})
	& = & \begin{cases}
			\delta_1 J_{20}(\psi_1, \psi_2)
				& \mbox{if } j = k = 1 , \\
			\delta_{k-1} J_{02}(\psi_{k-1},\psi_k) + \delta_k J_{20}(\psi_k,\psi_{k+1})
				& \mbox{if } 2 \le j = k < m , \\
			\delta_{m-1} J_{02}(\psi_{m-1},\psi_m)
				& \mbox{if } j = k = m , \\
			\delta_j J_{11}(\psi_j, \psi_k)
				& \mbox{if } 1 \le j = k-1 < m , \\
			0   & \mbox{if } |j - k| > 1 .
		\end{cases}
\end{eqnarray*}

\paragraph{Proof of \eqref{eq: coercivity of L}.}
In what follows let $\min(\bs{v})$ and $\max(\bs{v})$ denote the minimum and maximum, respectively, of all components of a vector $\bs{v}$. Moreover let $R(\bs{v}) := \max(\bs{v}) - \min(\bs{v})$. Then with $\bs{p} := (p_j)_{j=1}^m$ and $\bs{\delta} = (\delta_k)_{k=1}^{m-1}$, note first that
\begin{eqnarray*}
	L(\bs{\psi})
	& \le & \max(\bs{\psi}) - (x_m - x_1) \exp(\min(\bs{\psi})) \\
	& = & R(\bs{\psi}) + \min(\bs{\psi})
		- (x_m - x_1) \exp(\min(\bs{\psi})) \\
	& \to & - \infty	\quad\mbox{as } \|\bs{\psi}\| \to \infty \mbox{ while } R(\bs{\psi}) \le r_o
\end{eqnarray*}
for any fixed $r_o < \infty$. Secondly, let $\til{\psi}_j := \psi_j - \min(\bs{\psi})$. Then $\min(\til{\bs{\psi}}) = 0$, $\max(\til{\bs{\psi}}) = R(\bs{\psi})$, whence
\begin{eqnarray*}
	L(\bs{\psi})
	& = & \sum_{i=1}^m p_i \til{\psi}_i + \min(\bs{\psi})
		- \exp(\min(\bs{\psi})) \int_{x_1}^{x_m} \exp(\til{\psi}(x)) \, dx \\
	& \le & \left( 1 - \min(\bs{p}) \right) R(\bs{\psi})
		+ \sup_{s \in \R} \Bigl( s - \exp(s) \int_{x_1}^{x_m} \exp(\til{\psi}(x)) \, dx \Bigr) \\
	& = & \left( 1 - \min(\bs{p}) \right) R(\bs{\psi})
		- \log \int_{x_1}^{x_m} \exp(\til{\psi}(x)) \, dx - 1 \\
	& = & \left( 1 - \min(\bs{p}) \right) R(\bs{\psi})
		- \log \Bigl( \sum_{k=1}^{m-1} \delta_k J(\til{\psi}_k, \til{\psi}_{k+1}) \Bigr) - 1 \\
	& \le & \left( 1 - \min(\bs{p}) \right) R(\bs{\psi})
		- \log \Bigl( \min( \bs{\delta} ) J(0, R(\bs{\psi})) \Bigr) - 1 \\
	& = & \left( 1 - \min(\bs{p}) \right) R(\bs{\psi})
		- \log J(0, R(\bs{\psi})) - \log(e \min(\bs{\delta})) ,
\end{eqnarray*}
where we used the fact that $\max_{s \in \R} (s - \exp(s) A) = - \log A - 1$ for any $A > 0$. Moreover, for $r > 0$,
$$
	- \log J(0, r)
	\ = \ \log \Bigl( \frac{r}{e^r - 1} \Bigr)
	\ = \ - r + \log \Bigl( \frac{r}{1 - e^{-r}} \Bigr)
	\ \le \ - r + \log(1 + r) ,
$$
whence
$$
	L(\bs{\psi})
	\ \le \ - \min(\bs{p}) R(\bs{\psi}) + \log(1 + R(\bs{\psi})) - \log(e\min(\bs{\delta}))
	\ \to \ - \infty	\quad\mbox{as } R(\bs{\psi}) \to \infty .
	\eqno{\Box}
$$

\paragraph{Proof of Theorem~\ref{thm: 1dim functional}.}
It follows from strict concavity of $L$ and \eqref{eq: 1st dir deriv L} that the function $\psi$ equals $\check{\psi}$ if, and only if,
\begin{equation}
	\sum_{i=1}^m p_i v(x_i) \ = \ \int_{x_1}^{x_m} v(x) f(x) \, dx
	\label{eq: gradient condition}
\end{equation}
for any function $v \in \GG$.

Note that any vector $\bs{v} \in \R^m$ is a linear combination of the vectors $\bs{v}^{(1)}$, $\bs{v}^{(2)}$, \ldots, $\bs{v}^{(m)}$, where
$$
	\bs{v}_{}^{(k)} \ = \ \left( 1\{i \le k\} \right)_{i=1}^m .
$$
With the corresponding functions $v^{(k)} \in \GG$ we conclude that $\psi$ maximizes $L$ if, and only if,
\begin{equation}
	\sum_{i=1}^k p_i \ = \ \int_{x_1}^{x_m} v_{}^{(k)}(x) f(x) \, dx
	\label{eq: special gradient condition}
\end{equation}
for $1 \le k \le m$. Now the vector $\bs{v}^{(m)}$ corresponds to the constant function $v^{(m)} := 1$, so that \eqref{eq: special gradient condition} with $k = m$ is equivalent to $F(x_m) = 1$. In case of $1 \le k < m$,
$$
	v_{}^{(k)}(x) \ := \ \begin{cases}
		1                      & \mbox{if } x \le x_k , \\
		(x_{k+1} - x)/\delta_k & \mbox{if } x_k \le x \le x_{k+1} , \\
		0                      & \mbox{if } x \ge x_{k+1} ,
	\end{cases}
$$
and it follows from Fubini's theorem that
\begin{eqnarray*}
	\int_{x_1}^{x_m} v_{}^{(k)}(x) f(x) \, dx
	& = & \int_{x_1}^{x_m} \int_0^1 1\{u \le v_{}^{(k)}(x)\} \, du \, f(x) \, dx \\
	& = & \int_0^1 \int_{x_1}^{x_m} 1\{x \le x_{k+1} - u \delta_k\} f(x) \, dx \, du \\
	& = & \int_0^1 F(x_{k+1} - u \delta_k) \, du \\
	& = & \delta_k^{-1} \int_{x_k}^{x_{k+1}} F(r) \, dr .
\end{eqnarray*}
These considerations yield the characterization of the maximizer of $L$.

As for the first and second moments, equation~\eqref{eq: gradient condition} with $v(x) := x$ yields the assertion that $\sum_{i=1}^m p_i x_i$ equals $\int_{x_1}^{x_m} x f(x) \, dx$. Finally, let $\bs{v} := (x_i^2)_{i=1}^n$ and $v \in \GG$ the corresponding piecewise linear function. Then
\begin{eqnarray*}
	\sum_{i=1}^m p_i^{} x_i^2 - \int_{x_1}^{x_m} x^2 f(x) \, dx
	& = & \int_{x_1}^{x_m} (v(x) - x^2) f(x) \, dx \\
	& = & \sum_{k=1}^{m-1} \int_{x_k}^{x_{k+1}} (x - x_k)(x_{k+1} - x) f(x) \, dx \\
	& = & \sum_{k=1}^{m-1} \delta_k^3 J_{11}(\psi_k, \psi_{k+1}) .
\end{eqnarray*}\\[-9ex]
\strut	\hfill	$\Box$

\paragraph{Proof of Theorem~\ref{thm: KKstar and VVA}.}
It is well known from convex analysis that $\bs{\psi} \in \KK \cap \dom(L)$ belongs to $\KK_*$ if, and only if, $\bs{v}^\top \nabla L(\bs{\psi}) \le 0$ for any vector $\bs{v} \in \R^m$ such that $\bs{\psi} + t \bs{v} \in \KK$ for some $t > 0$. By the special form of $\KK$, the latter condition on $\bs{v}$ is equivalent to $\bs{v}_a^\top \bs{v} \ge 0$ for all $a \in A(\bs{\psi})$. In other words, $\bs{v} = \sum_{i=1}^m \lambda_i \bs{b}_i$ with $\lambda_a \ge 0$ for all $a \in A(\bs{\psi})$. Thus $\bs{\psi} \in \KK$ belongs to $\KK_*$ if, and only if, it satisfies \eqref{eq: KKstar}.

Similarly, a vector $\bs{\psi} \in \VV(A) \cap \dom(L)$ belongs to $\VV_*(A)$ if, and only if, $\bs{v}^\top \nabla L(\bs{\psi}) = 0$ for any vector $\bs{v}$ in the linear space
$$
	\bigl\{ \bs{v} \in \R^m : \bs{v}_a^\top \bs{v} = 0 \mbox{ for all } a \in A \bigr\}
	\ = \ \mathrm{span} \bigl\{ \bs{b}_i : i \in \{1,\ldots,m\} \setminus A \bigr\} .
$$
But this requirement is obviously equivalent to \eqref{eq: VVA}.	\hfill	$\Box$

\paragraph{Acknowledgements.}
This work was partially supported by the Swiss National Science Foundation. We are grateful to Charles Geyer for drawing our attention to active set methods and to Geurt Jongbloed for stimulating discussions about shape-constrained estimation.

\paragraph{Software.}
The methods of Rufibach~(2006, 2007) as well as the active set method from Section~\ref{Active Set} are available in the R package {\tt "logcondens"} written by K.\ Rufibach and L.\ D\"umbgen; see also D\"{u}mbgen and Rufibach (2011). Corresponding Matlab code is available from the first author's homepage on {\tt www.stat.unibe.ch}.

\end{document}